\documentclass[aps,prl,twocolumn,superscriptaddress]{revtex4}
\usepackage{amsmath, amssymb}
\usepackage{braket}
\usepackage{ulem}
\usepackage{soul}
\usepackage{dcolumn}
\usepackage{bm}
\usepackage{graphicx}
\usepackage{wasysym}
\usepackage{setspace}
\usepackage{mathrsfs}
\usepackage{color}
\usepackage{hyperref}
\usepackage{CJKutf8}
\hypersetup{colorlinks=true,linkcolor=blue,filecolor=blue,citecolor = blue,urlcolor=blue,}

\usepackage{tikz}
\usepackage{lipsum}
\begin{document}
\title{Exact Solution for Current-Driven Domain-Wall Dynamics Beyond Lorentz Contraction in Antiferromagnets with Dzyaloshinskii-Moriya Interaction}
\author{Mu-Kun Lee\footnote{Contact author: mukunlee01@gmail.com}}
\affiliation
{Department of Applied Physics, Waseda University, Okubo, Shinjuku-ku, Tokyo 169-8555, Japan}
\author{Rub\'{e}n M. Otxoa}
\affiliation{Hitachi Cambridge Laboratory, J. J. Thomson Avenue, Cambridge CB3 0HE, United Kingdom}
\author{Masahito Mochizuki\footnote{Contact author: masa\_mochizuki@waseda.jp}}
\affiliation
{Department of Applied Physics, Waseda University, Okubo, Shinjuku-ku, Tokyo 169-8555, Japan}
\begin{abstract}
We study current-driven domain-wall (DW) dynamics in antiferromagnets (AFMs) with Dzyaloshinskii-Moriya interaction (DMI). We obtain an exact analytical solution for spiral DW dynamics, applicable to both head-to-head DWs under bulk DMI and up-down DWs under interfacial DMI when the magnetic easy axis is aligned with the DMI vector. For the latter case experimentally relevant to synthetic AFMs with in-plane anisotropy, the solution predicts a constant DW velocity driven by nonadiabatic spin-transfer torque together with a steady rotation of the DW tilt angle induced by damping-like spin-orbit torque. Remarkably, the DW width shows unconventional current dependence, either pure elongation or contraction followed by elongation depending on damping and torque parameters, in sharp contrast to the Lorentz-type contraction known for antiferromagnetic (AF) DWs without DMI. These results provide an exact description of current-driven AF-DW dynamics and suggest experimentally accessible signatures of DMI-modified DW dynamics in synthetic AFMs.
\end{abstract}
\maketitle
\textit{Introduction.}---Solitons are localized solutions of nonlinear field equations characterized by finite energy and translationally invariant center coordinates, which endow them with particle-like properties~\cite{Zee,Lancaster}. In magnetic systems, domain walls (DWs) represent such topological solitons connecting oppositely magnetized domains~\cite{Tatara00,DW00} and can be driven by electric currents or magnetic fields, forming the basis of spintronic devices such as racetrack memory, logic, and neuromorphic computing architectures~\cite{racetrack,logic01,logic02,RC01,RC02,RC03}.

In recent years, antiferromagnetic (AF) DWs have attracted growing attention owing to their advantages over ferromagnets, including the absence of stray fields and ultrafast dynamics~\cite{AFM01,AFM02}. Their dynamics are governed by the N\'eel vector $\mathbf{n}=({\bf M}_{\rm A}-{\bf M}_{\rm B})/2$ with magnetization unit vector ${\bf M}_j$ for each sublattice $j = $ A, B. The Lagrangian of $\bf n$ contains a second-order time derivative. As a consequence, AF DWs exhibit relativistic dynamics analogous to field-theoretical kinks and undergo Lorentz-like boosts accompanied by a width contraction~\cite{Haldane,Kim,Shiino,Tatara,Beach,Otxoa01}.

In magnets lacking inversion symmetry, spin-orbit coupling induces the Dzyaloshinskii-Moriya interaction (DMI)~\cite{DM01,DM02}, which stabilizes spiral DWs~\cite{Tretiakov,Wang2} and other chiral magnetic textures such as skyrmions~\cite{sky01,sky02,sky03,sky04,Lee02}. However, when electric currents or magnetic fields drive these textures, their dynamics generally cannot be solved exactly. Instead, one must rely on approximate analytical treatments or numerical simulations~\cite{Thiele,Thiaville,Jung,Mougin,Yang2008,AF_WB,Lee01}. Exact solutions for current-driven magnetic textures therefore remain rare.

In this Letter, we show that the current-driven dynamics of DWs in AFMs with DMI admits an exact analytical solution under experimentally relevant conditions, namely synthetic AFMs with in-plane easy-axis anisotropy~\cite{Yang_Parkin,Yang_Parkin2,Duine,Bi,Saarikoski,Haltz_Mougin,Mallick,Wang,Hung_Ono}. The solution predicts an unconventional current dependence of the DW width, exhibiting either monotonic elongation or contraction followed by elongation depending on system parameters, in sharp contrast to the Lorentz contraction known for AF DWs without DMI. These results establish an exact theoretical framework for current-driven DW dynamics in AFMs with DMI and suggest experimentally accessible signatures of DMI-modified soliton dynamics.

%%%%%%%%%%%%%%%%%%%%%%%%%%%%%%%%%%%%%%%%%%%%%%%%%%
\begin{figure}[tb]
	%\centering
	\includegraphics[scale=0.4]{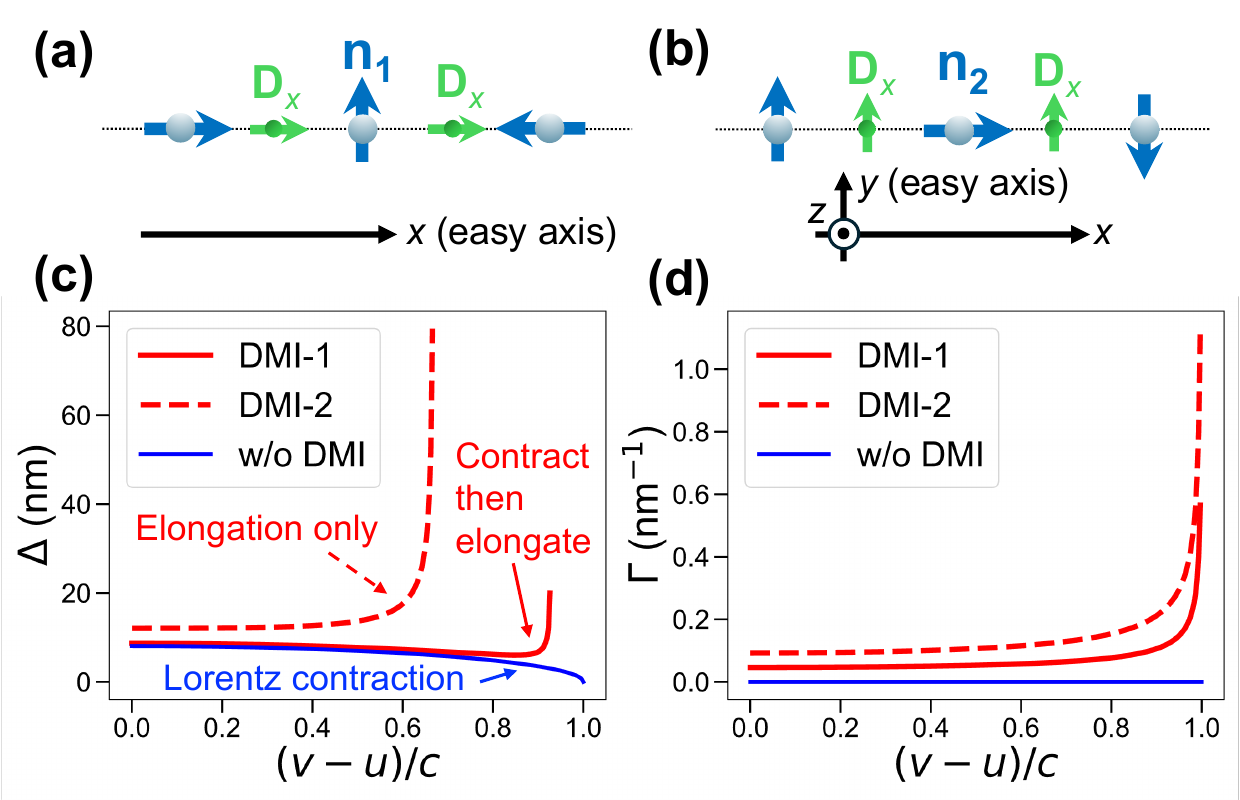}
	\caption{(a)--(b) Schematics of two types of DW profiles. Blue arrows indicate local N\'eel vectors, and green arrows show the DM vectors. (c) Domain wall width $\Delta$ and (d) tilt-angle wavevector $\Gamma$ versus $(v-u)/c$. Blue (red) curve indicates the case without (with) DMI. Parameters are $A=6.5$ pJ/m, $K=0.1$ MJ/m$^3$, and $D=1.2~ (0.6)$ mJ/m$^2$ for the red dashed (solid) curves, with $D^2-2AK>0$ ($<0$).}
	\label{Fig1}
\end{figure}
%%%%%%%%%%%%%%%%%%%%%%%%%%%%%%%%%%%%%%%%%%%%%%%%%%
\textit{Model.}---We first consider two types of one-dimensional systems and demonstrate their equivalence under an appropriate coordinate transformation. The first system consists of an AF texture with both the spatial variation and the easy axis oriented along the $\hat{\bf x}$ direction [e.g., a head-to-head DW shown in Fig.~\ref{Fig1}(a)] in the presence of bulk DMI~\cite{Tretiakov,Wang2}. The corresponding Hamiltonian density for the N\'eel vector is given by,
%%%%%%%%%%%%%%%%%%%%%%%%%%%%%%%%%%%%%%%%%%%%%%%%%%
\begin{align}
\mathcal{H}_1(\mathbf{n})=A(\partial_x\mathbf{n})^2-Kn^2_{x}-D(\mathbf{n}\cdot\bm{\nabla}\times\mathbf{n}),
\end{align}
%%%%%%%%%%%%%%%%%%%%%%%%%%%%%%%%%%%%%%%%%%%%%%%%%%
where $A$ is the intra-sublattice ferromagnetic (FM) exchange constant, $K$ is the easy-axis anisotropy energy density, and $D$ characterizes the strength of the bulk DMI. In the discrete representation, the DMI is written as $H_{\rm DMI}=\sum_{i,\mu}{\bf D}_\mu \cdot ({\bf n}_i \times {\bf n}_{i+\mu})$ with the DM vector ${\bf D}_\mu=\delta_{\mu x}D\hat{\bf x}$ ($\delta_{\mu\nu}$ denotes the Kronecker delta). This model describes, for example, an AF nanowire aligned along the direction of the DM vector.

The second system corresponds to AFMs with the easy axis oriented along the $\hat{\bf y}$ direction in the presence of interfacial DMI arising from broken inversion symmetry along the $\hat{\bf z}$ direction. Examples include noncentrosymmetric magnets such as K$_2$V$_3$O$_8$~\cite{Bogdanov,Velkov} and synthetic AFMs, which consist of two FM layers stacked along $\hat{\bf z}$ direction and coupled via the Ruderman-Kittel-Kasuya-Yosida interaction mediated by a nonmagnetic spacer layer~\cite{Yang_Parkin,Yang_Parkin2,Duine,Bi,Saarikoski,Haltz_Mougin,Mallick,Wang,Hung_Ono}. Assuming, for instance, an up-down DW configuration as shown in Fig.~\ref{Fig1}(b), the Hamiltonian density takes the form
%%%%%%%%%%%%%%%%%%%%%%%%%%%%%%%%%%%%%%%%%%%%%%%%%%
\begin{align}
\mathcal{H}_2(\mathbf{n})=A(\partial_x\mathbf{n})^2-Kn^2_y-D(n_x\partial_x n_z-n_z\partial_x n_x),
\label{H2}
\end{align}
%%%%%%%%%%%%%%%%%%%%%%%%%%%%%%%%%%%%%%%%%%%%%%%%%%
where $D$ now represents the strength of the interfacial DMI with the DM vector ${\bf D}_\mu=\delta_{\mu x}D\hat{\bf y}$.

Despite the different microscopic origins of the DMI, both Hamiltonians can be expressed in a unified form as
$\mathcal{H}=A(\partial_x{\bf n})^2-K({\bf n}\cdot\hat{\bf e}_{k})^2+D\hat{\bf e}_{k}\cdot{\bf n}\times\partial_x{\bf n}$,
where $\hat{\bf e}_k$ denotes the easy-axis direction. Introducing the parametrization ${\bf n}\cdot\hat{\bf e}_k=\cos\theta$, the anisotropy term takes an identical form regardless of the direction of $\hat{\bf e}_k$. The DMI can be written as $\sum_{ij}D\epsilon_{ijk}n_i\partial_xn_j$,where $\epsilon_{ijk}$ is the Levi-Civita symbol. Therefore, using the parametrizations $\mathbf{n}_1 \equiv (\cos\theta,\sin\theta\sin\phi,\sin\theta\cos\phi)$ for $\mathcal{H}_1$ and $\mathbf{n}_2 \equiv (\sin\theta\cos\phi,\cos\theta,\sin\theta\sin\phi)$ for $\mathcal{H}_2$, both Hamiltonians reduce to the identical form $\mathcal{H}_1(\mathbf{n}_1)=\mathcal{H}_2(\mathbf{n}_2)=A[(\theta')^2+\sin^2\theta(\phi')^2]-K\cos^2\theta-D\sin^2\theta(\phi')$, where the prime denotes differentiation with respect to $x$. Therefore, these two systems are mathematically equivalent and can be treated within a unified framework. In particular, the static spiral DW solution for $D^2<4AK$ coincides with the well-known exact solution in ferromagnets obtained by Tretiakov and Abanov, who analyzed $\mathcal{H}_1({\bf M})$ in terms of the magnetization ${\bf M}$~\cite{Tretiakov}. In the present work, we focus on the regime $D^2<4AK$.

\textit{Results.}---To obtain the dynamic soliton solution, we first consider the simplest case of an AF DW driven by the current-induced adiabatic spin-transfer torque (STT), which originates from the conservation of angular momentum between conduction electron spins and the local magnetization~\cite{Tatara00,Brataas}. In the exchange limit, where the AF exchange interaction is the dominant energy scale, the Lagrangian density can be written as~\cite{Tatara,Kim,AFM02,Haldane,Lee01,Lee02}
%%%%%%%%%%%%%%%%%%%%%%%%%%%%%%%%%%%%%%%%%%%%%%%%%%
\begin{align}
\mathcal{L}&=\frac{A}{c^2}(\dot{\mathbf{n}}+u\mathbf{n}')^2-\mathcal{H},
\label{L0}
\end{align}
%%%%%%%%%%%%%%%%%%%%%%%%%%%%%%%%%%%%%%%%%%%%%%%%%%
where $\dot{\mathbf{n}}$ denotes the time derivative of $\mathbf{n}$, $c$ is the maximum magnon group velocity, which plays the role of an effective speed of light in magnetic systems, and $u=p a_0^3 j_{\rm e}/(2e)$ is the spin-drift velocity associated with the electric current. Here, $j_{\rm e}$ is the current density, $p$ is the spin polarization of conduction electrons, $a_0$ is the lattice constant, and $e$ is the elementary charge. The adiabatic STT gives rise to the convective derivative term in the expression $(\dot{\bf n}+u{\bf n}')$ in Eq.~(\ref{L0})~\cite{Tatara00}.

We now consider the Hamiltonian $\mathcal{H}=\mathcal{H}_2({\bf n}_2)$ and assume a rigidly translating texture described by $\theta=\theta(x-vt)$ and $\phi=\phi(x-vt)$ with velocity $v$. This implies $\dot{\theta}=-v\theta'$ and $\dot{\phi}=-v\phi'$, and the corresponding Euler-Lagrange equations are
%%%%%%%%%%%%%%%%%%%%%%%%%%%%%%%%%%%%%%%%%%%%%%%%%%
\begin{align}
&A\gamma_v^2\sin\theta~\phi'' =\cos\theta~\theta'(D-2A\gamma_v^2\phi'),\nonumber\\
&2A\gamma_v^2\theta'' =\sin 2\theta(K-D\phi'+A\gamma_v^2\phi'^2),
\end{align}
%%%%%%%%%%%%%%%%%%%%%%%%%%%%%%%%%%%%%%%%%%%%%%%%%%
where $\gamma_v \equiv \sqrt{1-(v-u)^2/c^2}$ is the inverse Lorentz-like factor. This factor emerges naturally from the combined effect of the convective derivative in Eq.~(\ref{L0}) and the rigid-translation condition $\dot{\bf n}=-v{\bf n}'$. Substituting this relation into Eq.~(\ref{L0}) yields $A(\dot{\bf n}+u{\bf n}')^2/c^2=A\dot{\bf n}^2/\bar{c}^2$ with $\bar{c}=cv/(v-u)$. The Lagrangian then takes the form $\mathcal{L}=A\left[\dot{\bf n}^2/\bar{c}^2-({\bf n}')^2\right]-\mathcal{H}_{\rm DMI}-\mathcal{H}_{\rm ani}$, with the first term formally resembling the Lorentz-invariant field theory with an effective speed of light $\bar{c}$. This structure supports the Lorentz-like boost characterized by the factor $\gamma_v=\sqrt{1-v^2/\bar{c}^2}=\sqrt{1-(v-u)^2/c^2}$. Importantly, this apparent Lorentz invariance is only emergent and does not represent a fundamental symmetry of the system. The true characteristic velocity $c$ remains fixed, and the presence of STT explicitly breaks Lorentz invariance. The emergence of the factor $\gamma_v$ reflects instead the relative motion between the DW and the spin current. Consequently, it is the relative velocity $(v-u)$ between the DW and conduction electrons that governs the modulation of the DW profile.

When $\phi''=0$, we obtain an exact spiral DW solution,
%%%%%%%%%%%%%%%%%%%%%%%%%%%%%%%%%%%%%%%%%%%%%%%%%%
\begin{align}
&\phi=\Gamma(v)(x-vt)+\phi_0,\quad
\Gamma(v)=\frac{D}{2A\gamma_v^2},
\label{phi}
\\
&\theta=2\tan^{-1}\left[\exp\left(\frac{x-vt}{\Delta(v)}\right)\right],
\end{align}
%%%%%%%%%%%%%%%%%%%%%%%%%%%%%%%%%%%%%%%%%%%%%%%%%%
with a relativistic-like DW width given by
%%%%%%%%%%%%%%%%%%%%%%%%%%%%%%%%%%%%%%%%%%%%%%%%%%
\begin{align}
\Delta(v)=\gamma_v
\sqrt{\frac{A}{K-(D^2/4A\gamma_v^2)}}.
\label{delta_v}
\end{align}
%%%%%%%%%%%%%%%%%%%%%%%%%%%%%%%%%%%%%%%%%%%%%%%%%%
In the static limit with $v=u=0$ and $\gamma_v=1$, this solution reduces to the well-known Tretiakov-Abanov result for FM DWs with bulk DMI~\cite{Tretiakov}, yielding the renormalized DW width $\Delta(0)=\sqrt{A/(K-D^2/4A)}$ and a linear spatial variation of the tilt angle $\phi(x)=Dx/2A+\phi_0$. These results provide a direct experimental signature of interfacial DMI in synthetic AFMs with in-plane anisotropy, enabling quantitative determination of DMI strength from static DW profiles. Such characterization has remained largely unexplored, as most previous experimental and theoretical studies have focused on systems with perpendicular magnetic anisotropy~\cite{Yang_Parkin,Yang_Parkin2,Duine,Bi,Mallick,Hung_Ono}.

In the absence of DMI ($D=0$), the solution reduces to the conventional Lorentz-contracted DW width $\Delta=\gamma_v\sqrt{A/K}$, which is the well-known result for current-driven AF DWs~\cite{Tatara,Shiino,Otxoa01}. A striking effect of DMI is that, in addition to the overall Lorentz-like factor $\gamma_v$, additional $\gamma_v$ dependence appears inside the square root in Eq.~(\ref{delta_v}). As shown in Fig.~\ref{Fig1}(c), when the relative velocity $(v-u)/c\equiv \bar{v}$ increases from zero, the DW width exhibits two qualitatively distinct behaviors, that is,  (i) monotonic elongation (red dashed curve), and (ii) an initial contraction followed by rapid elongation (red solid curve). The criterion distinguishing these behaviors can be obtained from the $\bar{v}$-derivatives of  $\Delta$, that is, $d\Delta/d\bar{v}|_{\bar{v}=0}=0$ and $d^2\Delta/d\bar{v}^2|_{\bar{v}=0}=4A(D^2-2AK)/(4AK-D^2)^{3/2}$. When $D^2>2AK$, the curvature is positive, and the DW width increases monotonically with $\bar{v}$, corresponding to behavior (i). In contrast, when $D^2<2AK$, the curvature is negative, leading to an initial contraction followed by subsequent elongation, corresponding to behavior (ii). For the latter case (red solid curve), the denominator inside the square root in Eq.~(\ref{delta_v}) vanishes at a critical velocity [e.g., $\bar{v}\approx 0.93$ in Fig.~\ref{Fig1}(c)], indicating the loss of stability of the DW solution. Consequently, the DW width diverges, and the solution ceases to exist beyond this point. Furthermore, the tilt-angle wavevector $\Gamma$ also increases monotonically with $\bar{v}$ according to Eq.~(\ref{phi}), as shown in Fig.~\ref{Fig1}(d). These results demonstrate that DMI induces pronounced and experimentally observable modifications to the current-driven DW dynamics in AFMs, providing clear signatures of DMI in systems such as synthetic AFMs and bulk noncentrosymmetric AFMs when adiabatic STT is the dominant driving mechanism.

We now extend the above minimal model to a realistic synthetic AFM with in-plane magnetic anisotropy. In this system, the bottom heavy-metal layer injects a vertical spin current via the spin Hall effect~\cite{SHE01,SHE02}, which exerts damping-like spin-orbit torque (SOT) in the FM layers~\cite{SOT}. We assume that the damping-like SOT has equal magnitude in the two FM layers, which is a good approximation when the FM-layer thickness is smaller than the spin coherence length. For generality, we also include a sublattice-symmetric field-like SOT arising from the Rashba spin-orbit coupling in the FM layers due to the broken inversion symmetry of the multilayer structure~\cite{SOT02,SOT03}. This symmetry breaking also gives rise to interfacial DMI of equal strength in both FM layers~\cite{DMI01}. In addition, we incorporate sublattice-symmetric adiabatic and nonadiabatic STTs.

The dynamics of each sublattice magnetization ${\bf M}_j$ ($j$=A,B) obeys the Landau-Lifshitz-Gilbert-Slonczewski (LLGS) equation~\cite{LLGS01,LLGS02,LLGS03},
%%%%%%%%%%%%%%%%%%%%%%%%%%%%%%%%%%%%%%%%%%%%%%%%%%
\begin{align}
\dot{\mathbf{M}}_j=\gamma \mathbf{H}^{\rm eff}_j\times\mathbf{M}_j+\alpha\mathbf{M}_j\times \dot{\mathbf{M}}_j+\mathbf{T}_j,
\label{LLGS00a}
\end{align}
%%%%%%%%%%%%%%%%%%%%%%%%%%%%%%%%%%%%%%%%%%%%%%%%%%
where $\gamma$ and $\alpha$ are the gyromagnetic ratio and the Gilbert damping constant, respectively. The effective magnetic field is defined as  $\mathbf{H}^{\rm eff}_j=-(\delta  W/\delta\mathbf{M}_j)/(M_{\rm s}a^3_0)$ with the total magnetic energy given by $W=\int d^3x[\sum_{j}\mathcal{H}_2({\bf M}_j)+(J_{\rm AF}/a^3_0){\bf M}_{\rm A}\cdot{\bf M}_{\rm B}]$, where $J_{\rm AF}$ denotes the interlayer AF exchange coupling, and $M_{\rm s}$ is the saturation magnetization. The torque term $\mathbf{T}_j$ includes adiabatic and nonadiabatic STTs~\cite{STT01,STT02,STT03,STT04}, as well as damping-like and field-like SOTs, which is given by,
%%%%%%%%%%%%%%%%%%%%%%%%%%%%%%%%%%%%%%%%%%%%%%%%%%
\begin{align}
{\mathbf{T}}_j&=-u \mathbf{M}'_j+\beta u\mathbf{M}_j\times\mathbf{M}'_j
\notag\\
&+uc_{\rm H}\mathbf{M}_j\times \mathbf{M}_j\times \hat{\mathbf{y}}+uc_{\rm H}f_{\rm so}\mathbf{M}_j\times\hat{\mathbf{y}},
\label{LLGS00b}
\end{align}
%%%%%%%%%%%%%%%%%%%%%%%%%%%%%%%%%%%%%%%%%%%%%%%%%%
where $\beta$ is the nonadiabatic STT parameter. The coefficient $c_{\rm H}$ characterizes the strength of the spin Hall torque, which is given by $c_{\rm H}=\gamma \hbar \theta_{\rm SH}/(M_{\rm s}t_{\rm L} p a_0^3)$, where $\theta_{\rm SH}$ is the spin Hall angle, $t_{\rm L}$ is the FM-layer thickness, $\hbar$ is the reduced Planck constant, and $f_{\rm so}$ is the ratio between field-like and damping-like SOT components~\cite{SOT}.

In the strong AF exchange limit, the equation of motion for the N\'eel vector $\mathbf{n}$ can be derived from the LLGS equation as shown in our previous work~\cite{Lee02}, which is given by,
%%%%%%%%%%%%%%%%%%%%%%%%%%%%%%%%%%%%%%%%%%%%%%%%%%
\begin{align}
0&=\mathbf{n} \times \Big\{
\mathcal{D}^2\mathbf{n}+2\lambda_{\rm so}\hat{\bf y}\times\mathcal{D}\mathbf{n}
+\lambda^2_{\rm so}n_y\hat{\bf y}\Big\}
\notag \\
&+\frac{\gamma\bar{a}}{2}\mathbf{n}\times
\Big[\alpha \dot{\mathbf{n}}+\beta u\mathbf{n}'+uc_{\rm H}\mathbf{n}\times\hat{\bf y}
-\frac{\gamma}{2}\mathbf{f}_{\mathbf{n}}\Big],
\label{eq_n}
\end{align}
%%%%%%%%%%%%%%%%%%%%%%%%%%%%%%%%%%%%%%%%%%%%%%%%%%
where
%%%%%%%%%%%%%%%%%%%%%%%%%%%%%%%%%%%%%%%%%%%%%%%%%%
\begin{align}
&\mathbf{f}_{\mathbf{n}}=\frac{2}{M_{\rm s}}
\Big( A\mathbf{n}''+K n_y\hat{\bf y}+D(n'_z \hat{\bf x}-n'_x\hat{\bf z}) \Big),
\\
&\lambda_{\rm so}=\frac{uc_{\rm H}(f_{\rm so}-\alpha)}{1+\alpha^2},
\quad
\bar{a}=\cfrac{4J_{\rm AF}}{M_{\rm s}a_0^3}.
\end{align}
%%%%%%%%%%%%%%%%%%%%%%%%%%%%%%%%%%%%%%%%%%%%%%%%%%
Here $\bar{a}$ has the dimension of magnetic field (Tesla). The convective derivative is defined as $\mathcal{D} \equiv (\partial_t+\tilde{u}\partial_x)$ with the effective drift velocity $\tilde{u}=u(1+\alpha\beta)/(1+\alpha^2)$. This renormalized velocity arises from the interplay between STT and Gilbert damping in the Landau-Lifshitz formulation of the LLGS equation~\cite{Lee02}.

We find that Eq.~(\ref{eq_n}) admits an exact dynamic DW solution of the form $\theta(x,t)=-2\tan^{-1}{\exp[(x-q(t))/\Delta]}$ and $\phi(x,t)=\Gamma(x-q(t))+\phi_0(t)$, where $q(t)$ denotes the DW center and ${\bf n}=(\sin\theta\cos\phi, \cos\theta, \sin\theta\sin\phi)$. The solution yields closed expressions for the DW width $\Delta$, the tilt-angle wavevector $\Gamma$, the DW velocity $v=\dot q$, and the angular frequency $\omega=\dot{\phi}_0$ of the uniform tilt component [see Supplemental Material (SM)]:
%%%%%%%%%%%%%%%%%%%%%%%%%%%%%%%%%%%%%%%%%%%%%%%%%%
\begin{align}
&\Delta=\gamma_v\sqrt{A/F},
\label{delta}
\\
&F=K-\frac{D^2}{4A\gamma_v^2}
\notag \\
&+\frac{(1+\alpha f_{\rm so})[A(1+\alpha f_{\rm so}) c_{\rm H}+ D (\beta-\alpha)]u^2 c_{\rm H}}{u^2(\alpha-\beta)^2-\alpha^2(1+\alpha^2)^2\gamma^2 A\bar{a}/2M_{\rm s}},
\label{F}
\\
&\Gamma=\frac{1}{2A\gamma_v^2}\Big[D+\frac{4M_{\rm s}(\beta-\alpha)(1+\alpha f_{\rm so})}{\alpha^2(1+\alpha^2)^2\gamma^2\bar{a}}u^2 c_{\rm H}\Big].
\label{Gamma}
\end{align}
%%%%%%%%%%%%%%%%%%%%%%%%%%%%%%%%%%%%%%%%%%%%%%%%%%
Here $\gamma_v=\sqrt{1-(v-\tilde{u})^2/c^2}$ is the inverse Lorentz-like factor with $c=\gamma\sqrt{A\bar{a}/2M_{\rm s}}$, identical to the characteristic velocity obtained previously~\cite{Lee01}. This provides the first exact analytical solution for current-driven DW dynamics in AFMs with DMI and SOT. In the limit $c_{\rm H},\beta\to0$ and $\tilde{u}\approx u$ for small $\alpha$, these expressions reduce exactly to Eqs.~(\ref{phi}) and (\ref{delta_v}). Within the above DW ansatz, the collective-coordinate equations for $v$ and $\omega$ decouple as
%%%%%%%%%%%%%%%%%%%%%%%%%%%%%%%%%%%%%%%%%%%%%%%%%%
\begin{align}
\frac{2}{\gamma\bar{a}}
\left(\begin{array}{cc}
\dot{v} \\ 
\dot{\omega}
\end{array}\right)
=
\left(\begin{array}{cc}
-\alpha & 0 \\ 
0 & -\alpha
\end{array}\right)
\left(\begin{array}{cc} 
v \\ \omega 
\end{array}\right)
+
\left(\begin{array}{cc}
\beta u \\ 
-u c_{\rm H}
\end{array}\right),
\label{v_w}
\end{align}
%%%%%%%%%%%%%%%%%%%%%%%%%%%%%%%%%%%%%%%%%%%%%%%%%%
revealing particle-like DW dynamics with effective mass $2/\gamma\bar{a}$ and drag coefficient $\alpha$. 

In the steady state with $\dot v=\dot\omega=0$, we obtain 
%%%%%%%%%%%%%%%%%%%%%%%%%%%%%%%%%%%%%%%%%%%%%%%%%%
\begin{align}
v=\frac{\beta u}{\alpha}, \qquad
\omega=\frac{-u c_{\rm H}}{\alpha}
=\frac{-\gamma\hbar\theta_{\rm SH}}{2\alpha e M_{\rm s}t_{\rm L}}j_e.
\label{v_w2}
\end{align}
%%%%%%%%%%%%%%%%%%%%%%%%%%%%%%%%%%%%%%%%%%%%%%%%%%
Thus, while the DW velocity corresponds to the conventional spin-drift velocity multiplied by $\beta/\alpha$~\cite{Mougin,Yang2008}, the SOT generates constant rotation of the tilt angle, periodically transforming the DW between N\'eel and Bloch types with period $T=4\pi \alpha eM_{\rm s}t_{\rm L}/(\gamma\hbar\theta_{\rm SH}j_e)$ in the ps-ns range. This behavior contrasts with the Walker or Walker-breakdown dynamics in both FM DWs~\cite{Mougin,Yang2008} and AF DWs~\cite{AF_WB,Lee01}, where $v$ and $\omega$ become oscillatory above the breakdown threshold. Linearizing Eq.~(\ref{v_w}) around the steady state yields perturbations decaying as $\exp(-\alpha\gamma\bar{a}t/2)$, confirming the stability of the solution. The resulting dynamical DW configuration is given by $\mathbf{n}(x,t)=(n_x, n_y, n_z)$ where,
%%%%%%%%%%%%%%%%%%%%%%%%%%%%%%%%%%%%%%%%%%%%%%%%%%
\begin{align}
&n_x=\text{sech}[(x-vt)/\Delta]\cos[\Gamma(x-vt)+\omega t],
\nonumber
\\
&n_y=\tanh[(x-vt)/\Delta],
\nonumber
\\
&n_z=\text{sech}[(x-vt)/\Delta]\sin[\Gamma(x-vt)+\omega t].
\label{nz}
\end{align}
%%%%%%%%%%%%%%%%%%%%%%%%%%%%%%%%%%%%%%%%%%%%%%%%%%
For comparison, the coupled equations for the collective coordinates cannot in general be solved exactly and are usually treated using projection methods such as the Thiele approach~\cite{Thiele,Thiaville,Jung,Shiino} or approximate evaluations at the DW center~\cite{Lee01,Mougin}. Even for FM DWs with DMI, only the static Tretiakov-Abanov solution is exactly solvable, while current-driven dynamics requires perturbative treatments of the LLGS equation~\cite{Tretiakov}.

Notably, the present DW solution is nonchiral, as its handedness is not fixed by the sign of the DMI constant $D$.  For a general DW profile $\theta=2\zeta\tan^{-1}\exp[s(x-q)/\Delta]$ with $\zeta,s=\pm1$, the solutions for $v$, $\omega$, $\Delta$, and $\Gamma$ are identical for all $(\zeta,s)$.  Substituting the DW ansatz into the DMI energy $E_{\rm DMI}=-\int dx\,D(n_x\partial_x n_z-n_z\partial_x n_x)$ yields $E_{\rm DMI}=-2D\Gamma\Delta$, which is independent of $(\zeta,s)$ and constant in time. The energy becomes negative when $D\Gamma>0$, which can be satisfied irrespective of the sign of $D$ if the second term in the brackets of Eq.~(\ref{Gamma}) is small in the strong AF coupling and small-current regimes. On the other hand, in synthetic AFMs with perpendicular anisotropy, interfacial DMI prefers N\'eel-type DWs with chirality fixed by the sign of $D$~\cite{Yang_Parkin,Yang_Parkin2,Krishnia,Chen}.

The reason why damping-like SOT does not contribute to the DW velocity, while inducing tilt-angle rotation, can be intuitively understood from the AF exchange torque~\cite{Yang_Parkin,Yang_Parkin2,Lee01,Lee02}. Upon applying a current, the initial ${\bf M}_{\rm A,B}=\pm(\text{sech}(x)\cos\phi, \tanh(x), \text{sech}(x)\sin\phi)$ is rotated after an infinitesimal time $\delta t$ by the damping-like SOT to ${\tilde{\bf M}}_{j}= {\bf M}_j+(uc_{\rm H}\delta t)\,{\bf M}_j\times({\bf M}_j\times\hat{\bf y})+\mathcal{O}(\delta t^2)$. The AF coupling then induces a torque $\bm{\tau}_{\rm A,B}\propto -J_{\rm AF}\tilde{\bf M}_{\rm B,A}\times\tilde{\bf M}_{\rm A,B}\propto J_{\rm AF}(uc_{\rm H}\delta t)\hat{\bf y}\times{\bf M}_{\rm A,B}+\mathcal{O}(\delta t^2)$, which tends to rotate ${\bf M}_j$ around the $\hat{\bf y}$ axis and leads to $\omega\propto uc_{\rm H}$ in Eq.~(\ref{v_w2}). Since $\bm{\tau}_j$ has no component along the easy axis ($\hat{\bf y}$), it cannot drive the translational motion of DW.  When the easy axis lies along $\hat{\bf x}$ or $\hat{\bf z}$, Eq.~(\ref{eq_n}) no longer permits exact DW solutions. Instead, a Thiele-equation approach predicts a DW velocity dependent on SOT and nonlinear in current due to the nonlinear dependence of $\Delta$~\cite{Lee01,Collins,Shiino}.

%%%%%%%%%%%%%%%%%%%%%%%%%%%%%%%%%%%%%%%%%%%%%%%%%%
\begin{figure}[tb]
	\centering
	\includegraphics[scale=0.45]{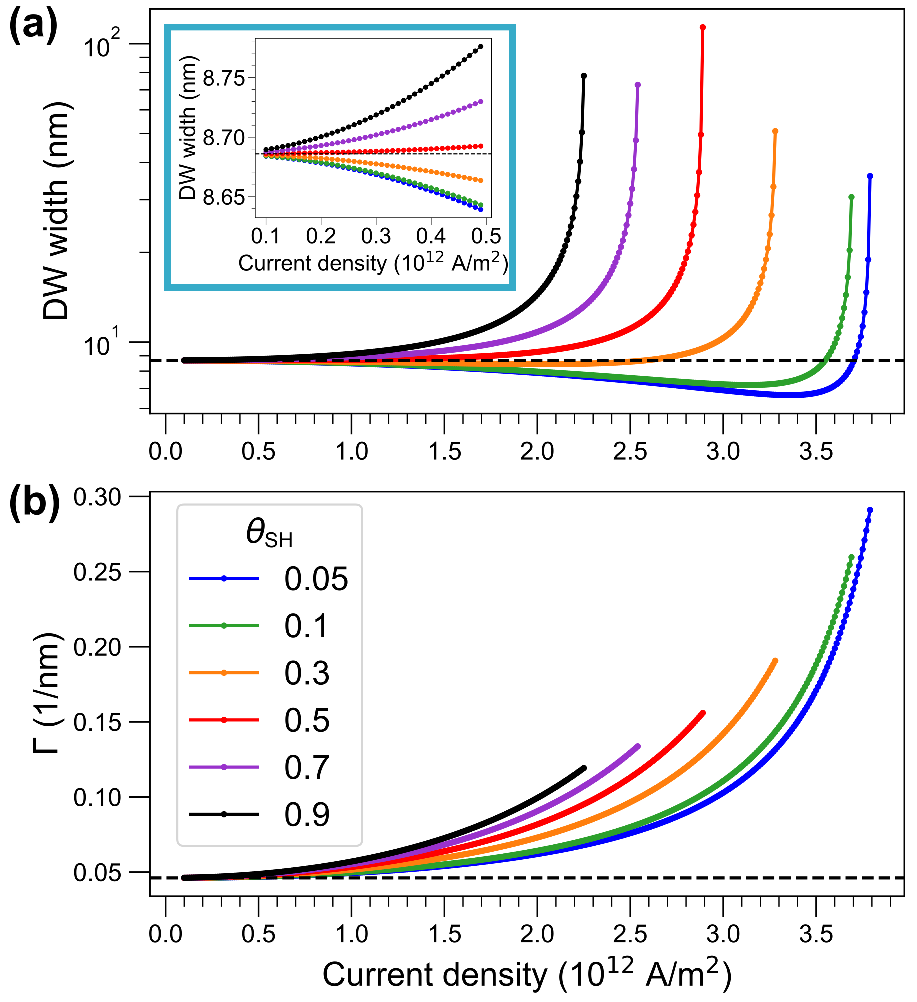}
	\caption{(a) DW width $\Delta$ and (b) tilt-angle wavevector $\Gamma$ versus current density for different values of $\theta_{\rm SH}$. The inset of (a) shows an enlarged view of the small-current regime. Black horizontal dashed lines indicate the values at zero current.}
	\label{Fig2}
\end{figure}
%%%%%%%%%%%%%%%%%%%%%%%%%%%%%%%%%%%%%%%%%%%%%%%%%%
Figure~\ref{Fig2} shows $\Delta$ and $\Gamma$ versus $j_{\rm e}$ for $\alpha=0.1$~\cite{Yang_Parkin}, $\beta=1.5\alpha$, $f_{\rm so}=0$, $a_0=1$~nm, $t_{\rm L}=10a_0$, $p=0.5$, $M_{\rm s}=0.5$~MA/m, $A=A_{\rm AF}=6.5$~pJ/m with $J_{\rm AF}=a_0A_{\rm AF}$, $K=0.1$~MJ/m$^3$, and $D=0.6$~mJ/m$^2$ ($D^2<2AK$ in this case). For various $\theta_{\rm SH}$, $\Gamma$ increases monotonically with current in Fig.~\ref{Fig2}(b), while $\Delta$ exhibits either monotonic elongation for $\theta_{\rm SH}\ge0.5$ or contraction followed by elongation for $\theta_{\rm SH}<0.5$, as seen in the inset of Fig.~\ref{Fig2}(a). For moderate currents $j_{\rm e}<3\times10^{12}$ A/m$^2$, $\Delta$ shows abrupt increase by nearly an order of magnitude. The curves terminate at the current density where $F$ in Eq.~(\ref{F}) changes from positive to negative, indicating the threshold current for the DW stability. 

%%%%%%%%%%%%%%%%%%%%%%%%%%%%%%%%%%%%%%%%%%%%%%%%%%
\begin{figure}[tb]
	\centering
	\includegraphics[scale=0.45]{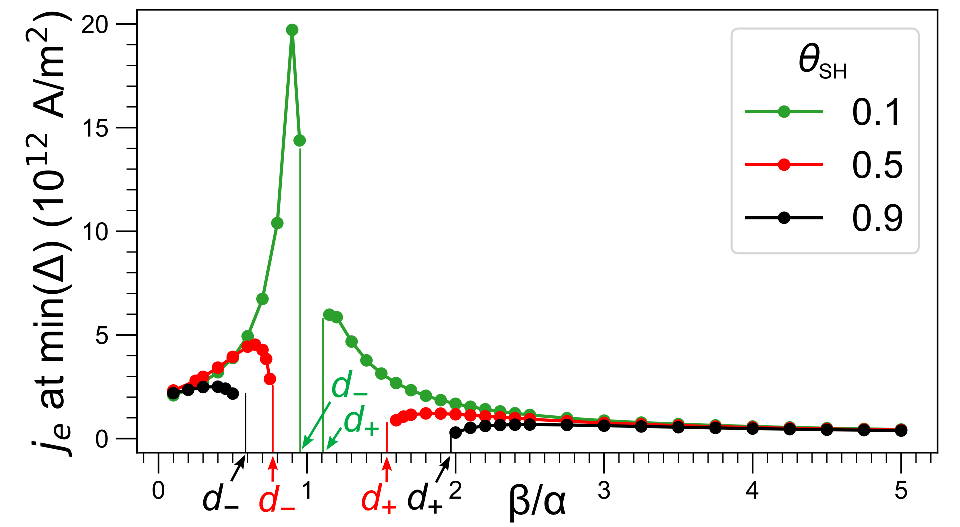}
	\caption{Current density $j_{\rm e}$ at which the DW width $\Delta$ becomes minimal for $D^2<2AK$. For each color corresponding to different $\theta_{\rm SH}$, $\Delta$ shows pure elongation when $d_-<\beta/\alpha<d_+$, while it first contracts and then elongates when $\beta/\alpha>d_+$ or $<d_-$.}
	\label{Fig3}
\end{figure}
%%%%%%%%%%%%%%%%%%%%%%%%%%%%%%%%%%%%%%%%%%%%%%%%%%
Figure~\ref{Fig3} shows numerical solutions of $j_{\rm e}$ at which $\Delta$ becomes minimal as a function of $\beta/\alpha$, marking the transition between contraction and elongation of $\Delta$. 
Using $d_{\pm}$ derived in the SM from the curvature of $\Delta$ at $j_{\rm e}=0$, when $D^2<2AK$, $\Delta$ contracts and then elongates for $\beta/\alpha>d_+$ or $\beta/\alpha<d_-$ for each $\theta_{\rm SH}$, whereas only elongation occurs for $d_-<\beta/\alpha<d_+$, consistent with the absence of numerical solutions for minimal $\Delta$ in this regime. When $4AK>D^2>2AK$, these two regimes of $\beta/\alpha$ interchange together with an exchange of $d_+\leftrightarrow d_-$.

\textit{Experimental relevance.}---The predicted DW dynamics should be experimentally accessible in synthetic AFMs where the two FM layers possess a slightly uncompensated net magnetization. In such systems, the N\'eel vector can be reconstructed by combining magnetic force microscopy (MFM) with scanning electron microscopy with polarization analysis (SEMPA)~\cite{Exp_SEMPA}. While the conventional DW-width contraction expected for AFMs without DMI is difficult to observe because it occurs on sub-$\mu$m length scales during ultrafast DW motion, the unconventional DW-width elongation predicted here can reach nearly an order-of-magnitude increase, providing a clear experimental signature. Another characteristic feature is the constant DW velocity predicted in our theory, in contrast to the nonlinear velocity saturation observed in ferrimagnetic systems~\cite{Beach}, which may be measured using magneto-optic Kerr microscopy (MOKE) with pulsed current injection~\cite{Yang_Parkin}. Even in compensated AFMs, the current-induced averaged magnetization is expected to exhibit characteristic oscillatory dynamics detectable by time-resolved X-ray magnetic circular dichroism (XMCD)~\cite{XMCD}. Details of this analysis and quantitative estimates are provided in SM. These considerations indicate a feasible route for experimental verification of the predicted DW dynamics.

\textit{Conclusion.}---We have investigated current-driven DW dynamics in AFMs with DMI. Starting from a general Lagrangian formulation including adiabatic STT, we revealed an unconventional DMI-induced modulation of the DW width, exhibiting either monotonic elongation or contraction followed by elongation with increasing current. Extending the analysis to synthetic AFMs with in-plane anisotropy, we obtained exact solutions for spiral DW dynamics characterized by unconventional width modulation, constant DW velocity driven by nonadiabatic STT, and steady tilt-angle rotation induced by damping-like SOT. These features provide experimentally accessible signatures and a predictive framework for current-driven soliton dynamics in antiferromagnets.

\textit{Acknowledgements.}---This work is supported by JSPS KAKENHI (No. 20H00337 and No. 25H00611), JST CREST (No. JP- MJCR20T1), and Waseda University Grant for Special Research Project (Grants No. 2024C-153, No. 2025C-133, and No. 2025C-134).

\clearpage
\onecolumngrid

\title{Supplemental Material: Exactly solvable current-driven domain wall dynamics in antiferromagnets with Dzyaloshinskii-Moriya interaction}

{\centering{\large\textbf{Supplemental Material: Exact Solution for Current-Driven Domain-Wall Dynamics Beyond Lorentz Contraction in Antiferromagnets with Dzyaloshinskii-Moriya Interaction}}}\\ \\
Mu-Kun Lee$^1$, Rub\'{e}n M. Otxoa$^2$, and Masahito Mochizuki$^1$\\
{\footnotesize \textit{$^1$Department of Applied Physics, Waseda University, Okubo, Shinjuku-ku, Tokyo 169-8555, Japan\\
$^2$Hitachi Cambridge Laboratory, J. J. Thomson Avenue, Cambridge CB3 0HE, United Kingdom}}

\tableofcontents
\section{Domain wall solution of staggered magnetization}
For a synthetic antiferromagnet (AFM)~\cite{Yang_Parkin,Duine} with sublattice-equivalent spin-orbit torques (SOTs) and spin-transfer torques (STTs), we have derived the equation of motion for staggered magnetization $\bf n$ from the Landau-Lifshitz form of LLGS equation in Ref.~\cite{Lee02} in the exchange limit with strong interlayer antiferromagnetic (AF) exchange interaction as Eq.~(10) in the main text, rewritten here as
\begin{eqnarray}
0&=&\mathbf{n}\times\mathcal{D}^2\mathbf{n}+\frac{\gamma\tilde{a}}{2}\mathbf{n}\times\Big[\alpha\dot{\mathbf{n}}+\beta u\mathbf{n}'+uc_{\rm H}\mathbf{n}\times\mathbf{y}-\frac{\gamma}{2}\mathbf{f}_{\mathbf{n}}\Big]
+\mathbf{n}\times\Big\{2\lambda_{\rm so}{\bf y}\times\mathcal{D}\mathbf{n}+\lambda_{\rm so}^2n_y{\bf y}\Big\}\equiv\mathbf{V},\label{eq_n_S}\\
\mathbf{f}_{\mathbf{n}}&=&\frac{2}{M_{\rm s}}\Big[{A}\mathbf{n}''+{K} n_y\mathbf{y}+{D}(n'_z\mathbf{x}-n'_x\mathbf{z})\Big].
\end{eqnarray}
We can roughly check the ratio of magnitudes of the terms in the curly brackets to other terms as follows. Suppose the magnetization is oscillating with a frequency $\omega_0$ and has a texture with a dependence $n_\mu=n_\mu((x-q)/\Delta)$ such that ${n}'_\mu\sim 1/\Delta$ and ${n}''_\mu\sim 1/\Delta^2$ ($\mu=x,y,z$), then the equation gives $\ddot{n}_\mu\sim \omega^2_0\sim (\gamma\tilde{a})[\mathcal{O}(\alpha\omega_0)+\mathcal{O}(\beta u/\Delta)+\mathcal{O}(u c_{\rm H})+\mathcal{O}(\gamma A/\Delta^2)+\mathcal{O}(K)+\mathcal{O}(D/\Delta)]$. For typical DWs, $\Delta/a_0\sim 10^1$.
If we consider $\alpha\omega_0$ as the largest frequency scale in the bracket, then $\omega_0\sim\alpha \gamma\tilde{a}$. 
Since $\lambda_{\rm so}=uc_{\rm H}(\alpha-f_{\rm so})/(1+\alpha^2)$ and both $\alpha$ and $f_{\rm so}$ are assumed much smaller than one, the terms in $\{...\}$ of the above equation are then much smaller compared to the $c_{\rm H}$ term in $[...]$. 

To solve $\bf n$, we define ${\bf n}=(\sin\theta\cos\phi, \cos\theta, \sin\theta\sin\phi)$, and assume the dynamically steady motion of DW by taking $\theta(x,t)=\theta(x-q(t))$, $\phi(x,t)=\Gamma(x-q(t))+\phi_0(t)$. With the presumed form $\theta=-2\tan^{-1}[\exp(\frac{x-q(t)}{\Delta})]$, we can get two independent equations by taking $0=(\mathbf{V}\cdot\mathbf{x})\sin\phi-(\mathbf{V}\cdot\mathbf{z})\cos\phi$ and $0=(\mathbf{V}\cdot\mathbf{y})$ with $\bf V$ defined in Eq.~(\ref{eq_n_S}) as
\begin{eqnarray}
0&=&\sin(2\theta)\Big[\frac{2\gamma^2\bar{a}A/M_{\rm s}-4(v-\tilde{u})^2}{\Delta^2}-\frac{2\gamma^2\bar{a}}{M_{\rm s}}[K+\Gamma(\Gamma A-D)]+4[\Gamma(\tilde{u}-v)+\omega]^2
-8\lambda_{\rm so}[\Gamma(\tilde{u}-v)+\omega]+4\lambda_{\rm so}^2\Big]\label{Eq1}\\
&+&\frac{4\sin\theta}{\Delta}\Big[\gamma\bar{a}(\alpha v-\beta u)+2\dot{v}\Big]\label{Eq2},\\
0&=&\cos\theta\Big[8\Gamma(v-\tilde{u})^2+\frac{2\gamma^2\bar{a}}{M_{\rm s}}(D-2\Gamma A)-8(v-\tilde{u})(\omega-\lambda_{\rm so})\Big]\label{Eq3}\\
&+&2\Delta\Big[\gamma\bar{a}[u c_{\rm H}-\Gamma(\alpha v-\beta u)+\alpha\omega]-2(\Gamma\dot{v}-\dot{\omega})\Big].\label{Eq4}
\end{eqnarray}
Since $\theta(x-q(t))$ depends on $x$, actually we have four locally independent equations as the requirements of vanishing for coefficients of $\sin(2\theta),\sin\theta,\cos\theta$, and the constant term on the right-hand sides of these two equations, which can be solved to give the unknown $\Delta, \Gamma$, $v\equiv\dot{q}$, and $\omega\equiv\dot{\phi}_0$. 
The coefficients of $\sin(2\theta)$ in Eq.~(\ref{Eq1}) and $\cos\theta$ in Eq.~(\ref{Eq3}) give 
\begin{eqnarray}
\Delta&=&\gamma_v\sqrt{A/F},\label{delta_S}\\
\Gamma&=&\frac{1}{2A\gamma^2_v}\Big[D-\frac{4M_{\rm s}(v-\tilde{u})(\omega-\lambda_{\rm so})}{\gamma^2\bar{a}}\Big],\label{Gamma_S}\\
F&=&K-\Gamma(D-\Gamma A)-\frac{2M_{\rm s}[\Gamma(v-\tilde{u})-\omega+\lambda_{\rm so}]^2}{\gamma^2\bar{a}},\label{F_S}
\end{eqnarray}
with $\gamma_v=\sqrt{1-(v-\tilde{u})^2/c^2}$ and $c=\gamma\sqrt{A\bar{a}/2M_{\rm s}}$.
The vanishing of the other two coefficients in Eqs.~(\ref{Eq2}) and (\ref{Eq4}) gives Eq.~(16) in the main text. Substituting the steady solutions of $v$ and $\omega$ in Eq.~(17) of main text into above Eqs.~(\ref{delta_S})--(\ref{F_S}), we get Eqs.~(13)--(15) in the main text. 
\section{Criterion for two types of DW width modulation}
In this section, we derive an analytical formula for the model parameters in the realistic synthetic AFM case that sets a criterion to distinguish whether the DW width will only elongate or will first contract then elongate as the current increases from zero. 
From Eq.~(13) in the main text, after some algebra we can find that $d\Delta(u)/du|_{u=0}=0$, which means the slope of $\Delta(u)$ is zero at zero current. Therefore, we need to calculate the sign of second derivative $d^2\Delta(u)/du^2|_{u=0}$ to check whether the slope $d\Delta(u)/du|_{u=0}$ concaves upwards or downwards at an infinitesimal $u$, with the former or later situation indicating that the DW width will first elongate or contract when $u$ increases from zero, respectively. Interestingly, one can find $d^2\Delta/du^2|_{u=0}$ is quadratic in $\beta/\alpha$ and has the form
\begin{eqnarray}
&&\frac{d^2\Delta}{du^2}\Big|_{u=0}\propto-(2AK-D^2)\Big(\frac{\beta}{\alpha}-d_{-}\Big)\Big(\frac{\beta}{\alpha}-d_{+}\Big),\label{cri}\\
&&d_{\pm}=1+\frac{Ac_{\rm H}\Big(D\pm\sqrt{4AK-D^2}\Big)(1+\alpha f_{\rm so})}{\alpha(2AK-D^2)}.\label{d}
\end{eqnarray}
The other multiplication factors in Eq.~(\ref{cri}) are all positive thus we did not show them in the equation. In this work we consider $4AK>D^2$ to stabilize the spiral DW at static state as found in the Tretiakov-Abanov solution. In the case of $2AK>D^2$, the above equations indicate the two distinct behaviors of $\Delta$ as:
\begin{eqnarray}
&&\text{(i) }\frac{\beta}{\alpha}>d_{+}\text{ or }\frac{\beta}{\alpha}<d_{-}\Rightarrow \Delta\text{ contracts at infinitesimal }u,\\
&&\text{(ii) }d_{-}<\frac{\beta}{\alpha}<d_{+}\Rightarrow \Delta\text{ elongates at infinitesimal }u.
\end{eqnarray}
In the other case of $4AK>D^2>2AK$, the above behaviors of $\Delta$ in (i) and (ii) switch, together with an exchange of $d_{+}\leftrightarrow d_{-}$ since $d_{\pm}$ in Eq.~(\ref{d}) also depends on the sign of $(2AK-D^2)$. 
Although there is no simple solution of $u$ at which $d\Delta/du=0$ that enables one to find the turning points of $\Delta$, this analysis of the concaving behaviors at $u=0$ turns out to give us the criterion of $\beta/\alpha$ that distinguishes whether $\Delta$ will contract then elongate or will only elongate as current increases, as shown in Fig.~3 in the main text.
\section{Averaged magnetization and its possible detection by XMCD}
The averaged magnetization ${\bf m}=({\bf M}_{\rm A}+{\bf M}_{\rm B})/2$ is a slave variable controlled by the dynamics of $\bf n$ in the exchange limit with large interlayer AF exchange interaction. It can be nonzero even in compensated AFMs.
Since in AFMs the detection of $\bf n$ is challenging, here we calculate the dynamics of $\bf m$ for the purpose of experimental observation of our DW solution. As derived in our previous work~\cite{Lee02}, the dominant part of $\bf m$ in the zeroth orders of Gilbert damping, external magnetic field, and field-like SOT, has the form
\begin{eqnarray}
{\bf m}&\approx&-\frac{2}{\bar{a}\gamma}{\bf n}\times(\partial_t+\tilde{u}\partial_x){\bf n}.
\end{eqnarray}
Substituting $\bf n$ in Eq.~(18) of the main text into this equation gives 
\begin{eqnarray}
{\bf m}&=&\frac{2}{\bar{a}\gamma}\text{sech}\Big(\frac{x-vt}{\Delta}\Big)\nonumber\\
&\times&\Big(-\frac{v-\tilde{u}}{\Delta}\sin[\Gamma(x-vt)+\omega t]+[\omega-\Gamma(v-\tilde{u})]\cos[\Gamma(x-vt)+\omega t]\tanh\Big(\frac{x-vt}{\Delta}\Big),\nonumber\\
&&\text{sech}\Big(\frac{x-vt}{\Delta}\Big)[\omega-\Gamma(v-\tilde{u})],\nonumber\\
&&\frac{v-\tilde{u}}{\Delta}\cos[\Gamma(x-vt)+\omega t]-[\omega-\Gamma(v-\tilde{u})]\sin[\Gamma(x-vt)+\omega t]\tanh\Big(\frac{x-vt}{\Delta}\Big)\Big).
\end{eqnarray}
Since it is challenging to experimentally detect the local profile of $\bf m$ due to the DW motion by, e.g., Magneto-optic Kerr effect (MOKE), here we calculate the spatial average of $\bf m$ through the quasi-one-dimensional synthetic AFM, $\int^\infty_{-\infty} m_i dx$, which may be detected by e.g., time-resolved X-ray magnetic circular dichroism (Tr-XMCD). Define $L$ as the sample length, we obtain
\begin{eqnarray}
\langle m_x\rangle=\frac{1}{L}\int m_x dx&=&\frac{2}{\bar{a}\gamma L}\int dx\Big\{\text{sech}\Big(\frac{x-vt}{\Delta}\Big)\frac{-v+\tilde{u}}{\Delta}\Big(\sin[\Gamma(x-vt)]\cos\omega t+\cos[\Gamma(x-vt)]\sin\omega t\Big)\nonumber\\
&+&\text{sech}\Big(\frac{x-vt}{\Delta}\Big)\tanh\Big(\frac{x-vt}{\Delta}\Big)[\omega-\Gamma(v-\tilde{u})]\Big(\cos[\Gamma(x-vt)]\cos\omega t-\sin[\Gamma(x-vt)]\sin\omega t\Big)\Big\}\nonumber\\
&=&-\frac{2\pi}{\bar{a}\gamma L}\Big\{v-\tilde{u}+\Gamma\Delta^2[\omega-\Gamma(v-\tilde{u})]\Big\}\text{ sech}\Big(\frac{\pi\Gamma\Delta}{2}\Big)\sin\omega t\equiv \langle m_{x0}\rangle \sin\omega t,\\
\langle m_y\rangle=\frac{1}{L}\int m_y dx&=&\frac{4\Delta}{\bar{a}\gamma L}[\omega-\Gamma(v-\tilde{u})],\\
\langle m_z\rangle=\frac{1}{L}\int m_z dx&=&\frac{2\pi}{\bar{a}\gamma L}
\Big\{v-\tilde{u}-\Gamma\Delta^2[\omega-\Gamma(v-\tilde{u})]\Big\}\text{ sech}\Big(\frac{\pi\Gamma\Delta}{2}\Big)\cos\omega t\equiv \langle m_{z0}\rangle \cos\omega t.
\end{eqnarray}
While $\langle m_y\rangle$ is time-independent, $\langle m_x\rangle$ and $\langle m_z\rangle$ show oscillations in time proportional to $\sin\omega t$ and $\cos\omega t$, respectively. 
Since the period $2\pi/\omega$ is in the ps--ns timescale and is inversely proportional to current, the oscillations of $\langle m_x\rangle, \langle m_z\rangle$ and static behavior of $\langle m_y\rangle$ may be detectable in synthetic AFMs by Tr-XMCD, which can achieve a time resolution under 50~ps with a moderate X-ray beam size of $\sim 10^4$~$\mu$m$^2$ to marginally cover the whole quasi-one-dimensional system~\cite{XMCD}. 
In order to measure these data, the static DW width is preferable to be smaller than $\mu$m scale, such that the DW center will remain in the bulk sample with length $L\sim 10^2 \mu$m during its motion, to fulfill the validity of the above spatial integrals that have relied on a change of variable as $x-vt\rightarrow x'$ followed by integration over $x'$.
In Supplementary~Fig.~\ref{Fig1_S}, we show the trends of amplitudes of $\langle m_i\rangle$ versus current taking the same model parameters for Fig.~2 in the main text with $\theta_{\rm SH}=0.5$, $\beta=2\alpha$ and $0.4\alpha$ in (a) and (b), respectively. The behaviors of $\langle m_{x0}\rangle$ and $\langle m_{z0}\rangle$ switches when $\beta/\alpha$ becomes from larger to less than 1, while $\langle m_y\rangle$ always shows a monotonic decrease with current. 
These trends of $\langle m_i\rangle$, together with DW velocity $v$ measured by MOKE using pulsed currents~\cite{Yang_Parkin}, and $\omega$ extracted from temporal oscillations of $\langle m_{x,z}\rangle$, may be used to indirectly infer the dependence of $\Delta$ and $\Gamma$ on current. Note that $\bar{a}$ can be tuned in synthetic AFMs by varying the thickness of the intermediate nonmagnetic layer. Reducing $\bar{a}$ while maintaining the approximate dynamics in exchange limit enhances the amplitude of $\bf m$, thereby improving the potential feasibility of experimental detection.
\begin{figure}[h]
	\centering
	\includegraphics[scale=0.6]{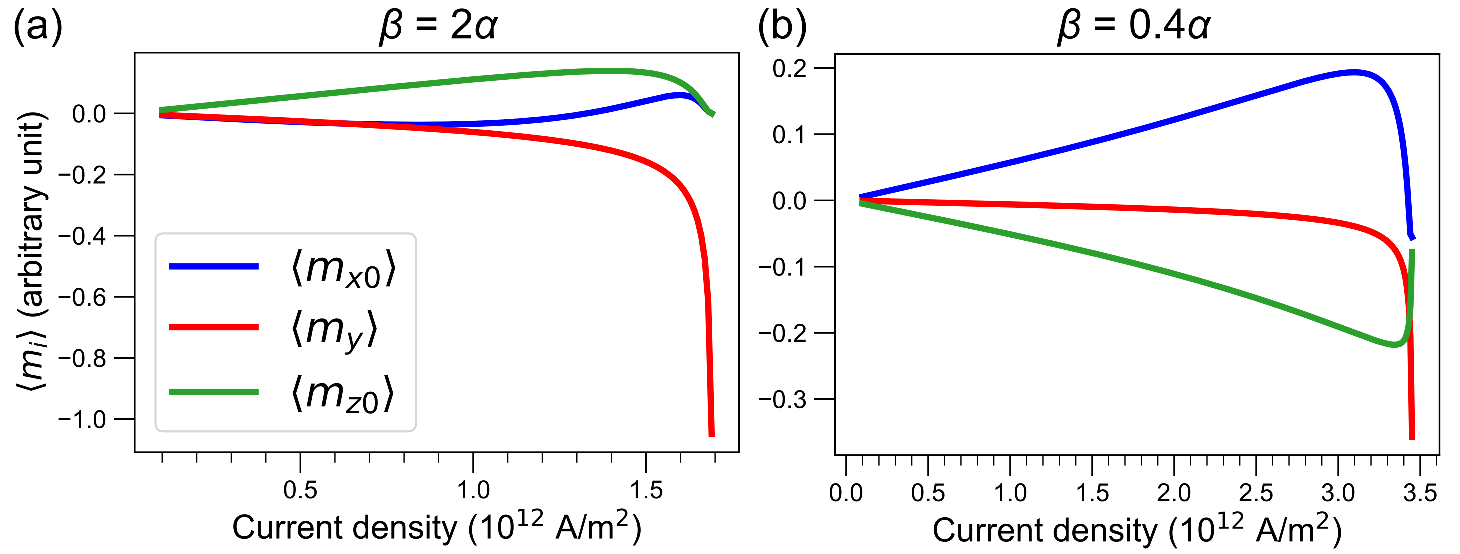}
	\renewcommand{\figurename}{Supplementary Fig.}\setcounter{figure}{0}
	\caption{Trends of amplitudes of $\langle m_i\rangle$ versus current for (a) $\beta=2\alpha$ and (b) $\beta=0.4\alpha$.}
	\label{Fig1_S}
\end{figure}

\end{document}